\documentclass{aastex631}

\usepackage{orcidlink}

\shorttitle{Decoding rotating neutron stars}
\shortauthors{L. L. Lopes\orcidlink{0000-0003-0982-9774}}

\graphicspath{{./}{figures/}}

\begin{document}

\title{Decoding rotating neutron stars: Role of the symmetry energy slope}



\author{Luiz L. Lopes}
\affiliation{Centro Federal de Educa\c{c}\~ao Tecnol\'ogica de Minas Gerais; \\ Campus VIII  CEP 37.022-560, Varginha - MG - Brasil
\email{llopes@cefetmg.com}
}



\begin{abstract}

In December 2023, the Fermi LAT Catalog announced the discovery of 33 new millisecond pulsars. Motivated by that, in this work, I study how different values of the symmetry energy slope $(L)$ affect the properties of static and slowly rotating neutron stars. For fixed values of angular velocity, I study how the slope influences the increase of the maximum mass,  the radii of the canonical 1.4 solar mass, its eccentricity, as well the same quantities for the 2.01 $M_\odot$ stars. I show that different slope values cause different variations not only the absolute quantities but also in relative ones. Indeed, different slope values predict different values for the eccentricity, which does not depend on the absolute value of the neutron stars' radii. Therefore, this quantity can be a powerful tool to constrain the symmetry energy slope. 
\end{abstract}



\section{Introduction} \label{sec:intro}

 Pulsars are among the fastest-spinning objects in nature, and their period can be as low as a few milliseconds. The first  millisecond pulsar was discovered over forty years ago, the PSR B1937+21, with a frequency of 642 Hz ($\Omega$ = 4034 $s^{-1}$), ~\citep{Backer1982}. Currently, PSR J1748-2446ad is the fastest-spinning pulsar known, at 716 Hz ($\Omega$ = 4499 $s^{-1}$)~\citep{Hessels2006}. Another remarkable example is the black widow pulsar PSR J0952-0607, which is not only the second fastest-spinning pulsar, rotating at 707 Hz ($\Omega$ = 4442 $s^{-1}$), but also it is potentially the heaviest neutron star ever known, whose mass lies between 2.35 $\pm$ 0.17 $M_\odot$~\citep{Romani_2022}. A better understanding of the effects of rotation in neutron stars become even more important in the last weeks. In December 2023, the third Fermi Large Area Telescope Catalog of gamma-ray pulsars announced the discovery of 33 new millisecond pulsars, among these, at least a dozen had an inferred angular velocity above 3500 $s^{-1}$~\citep{Smith2023}.

The modern theory of pulsars considers them as rotating neutron stars.  In the neutron star's crust, where the densities are usually below half of the saturation density, the matter is inhomogeneous and presents a high degree of complexity. In this regime clusters coexisting with neutron (super)fluid. Moreover, the competition between attractive nuclear force and repulsive Coulomb interaction can turn the nuclear matter into a frustrated system, i.e., the system presents more than one low-energy configuration, which
can cause the onset of unusual nuclear shapes with different geometries. This is called nuclear pasta phase~\citep{Ravenhall_1983,Lorenz1993}. Although these properties are relevant for rotating neutron stars since they are believed to affect pulsar glitches, their studies are beyond the scope of the present work. I refer the interested reader to~\citet{Chamel_2008} and the references therein.   On the other hand, in the neutron stars' core, where the densities can reach several times the nuclear saturation density, pressure is so strong that  nucleons can no longer form clusterized matter, and neutrons, protons, and leptons form a homogeneous liquid in beta-equilibrium.  The tool to describe a neutron star's interior is therefore many-body nuclear physics. A good model must be able to reproduce at least five well-constrained parameters at the saturation density: the nucleon effective mass ($M_N^*/M_N$), the binding energy per baryon ($E/A$), the compressibility ($K$), the symmetry energy ($S_0$), and the saturation density itself ($n_0$). 

A sixth important parameter is the symmetry energy slope, or simply slope $(L)$, However, unlike the other five parameters, the slope is not yet well constrained. In the earlier 2010s, most studies pointed to a relatively low value for $L$. For instance, in refs.~\citet{Paar2014,Steiner2014,Lattimer2013}  upper limits of 54.6, 61.9, and 66 MeV respectively were suggested. However, in the last couple of years, the situation has changed and new experiments have pointed to a significantly higher upper limit.
For instance, in a study about the spectra of pions in intermediate energy collisions, an upper limit of 117 MeV was obtained~\citep{pions}, while in one of the PREX II analyses~\citep{PREX2} an upper limit of 143 MeV was suggested.
All these conflicting results have been well summarized in a recent paper~\citep{Tagami2022}: the CREX group points to a slope in the range 0 $<~L~<$ 51 MeV, while PREX II results point to 76 MeV $<~L~<$ 165 MeV. The CREX and PREX II results do not overlap. That is a huge problem that must be solved.

It is also well-known that the symmetry energy slope affects the neutron star properties~\citep{Steiner2014,Lattimer2013,Rafa2011,Lopes2014BJP}. Therefore,  neutron star observations can indeed be used to constrain the symmetry energy slope.  However, most works on this subject use the static approximation. As neutron stars can spin very fast, the effects of their rotation can influence their macroscopic properties. Quantifying the effects of the slope in rotating neutron stars is therefore our main goal.

In this work, I study how different values of the symmetry energy slope affect neutron stars that rotate at different angular velocities. To accomplish this task, I use quantum hadrodynamics (QHD) with the traditional $\sigma-\omega-\rho$ mesons~\citep{Glenbook,Serot_1992}. Moreover, to keep the symmetry energy fixed while varying the slope, we use two extensions of QHD: to reduce the slope, we add the non-linear $\omega-\rho$ coupling as presented in the IUFSU model~\citep{IUFSU,Rafa2011,dex19jpg}, while to increase the slope, we add the scalar-isovector $\delta$ meson~\citep{KUBIS1997,Liu2002}.
The rotation is introduced via Hartle and Thorne's slowly-rotating formalism~\citep{Hartle_1967,Hartle_1968,Hartle_1973}. In general, most of the studies related to rotating neutron stars compare the static results with those in which the star rotates at the mass shedding limit, also called Keplerian limit, $\Omega_K$~\citep{Weber1992,Chu2000,Dhiman_2007,Jha_2008,Haensel_2009,Most_2020,Rather_2021}. Here, I try a more realistic approach, and study rotating neutron stars with angular velocity close to those observed in nature.
As pointed out earlier, the fastest-spinning pulsar has an angular velocity of $\Omega$ = 4499 $s^{-1}$. As even fast-spinning neutron stars may exist,  I use $\Omega$ = 5000 $s^{-1}$, as the maximum angular velocity, which corresponds to around 10$\%$ above the angular velocity of the PSR J1748-2446ad. This implies that most of the neutron stars studied here are significantly below the mass shedding limit, which guarantees Hartle and Thorne's slowly rotating approach as a good approximation.

\section{Nuclear model}
\label{sec:nuclear_model}

To describe the nuclear interaction, I use here an extended version of the quantum hadrodynamics (QHD)~\citep{Glenbook,Serot_1992}, which includes the $\omega-\rho$ non-linear coupling~\citep{IUFSU,Rafa2011,dex19jpg}, as well the scalar-isovector $\delta$ meson~\citep{KUBIS1997,Liu2002}. Considering a pure nucleonic matter, the Lagrangian density reads:
\begin{eqnarray}
\mathcal{L}_{QHD} =  \bar{\psi}_N \big[\gamma^\mu(\mbox{i}\partial_\mu  - g_{\omega}\omega_\mu   - g_{\rho} \frac{1}{2}\vec{\tau} \cdot \vec{\rho}_\mu)  \nonumber 
 - (M_N - g_{\sigma}\sigma - g_{\delta}\vec{\tau} \cdot \vec{\delta}) \big]\psi_N     \\
  + \frac{1}{2}(\partial_\mu \sigma \partial^\mu \sigma - m_s^2\sigma^2)  
 + \frac{1}{2}(\partial_\mu \vec{\delta} \cdot \partial^\mu \vec{\delta} - m_\delta^2\delta^2)  
 - \frac{1}{4}\Omega^{\mu \nu}\Omega_{\mu \nu} \nonumber  + \frac{1}{2} m_v^2 \omega_\mu \omega^\mu   \\
 + \Lambda_{\omega\rho}(g_{\rho}^2 \vec{\rho^\mu} \cdot \vec{\rho_\mu}) (g_{\omega}^2 \omega^\mu \omega_\mu) \nonumber   - U(\sigma) 
 + \frac{1}{2} m_\rho^2 \vec{\rho}_\mu \cdot \vec{\rho}^{ \; \mu} - \frac{1}{4}\bf{P}^{\mu \nu} \cdot \bf{P}_{\mu \nu} . \label{s1} 
\end{eqnarray}
Here, $\psi_N$ is the Dirac field of the nucleons, and $\sigma$, $\omega_\mu$, $\vec{\delta}$, and $\vec{\rho}_\mu$ are the mesonic fields. The $g$'s are the Yukawa coupling constants that simulate the strong interaction, $M_N$ is the nucleon mass, and $m_s$, $m_v$, $m_\delta$, and $m_\rho$ are the masses of the $\sigma$, $\omega$, $\delta$, and $\rho$ mesons, respectively. The self-interaction term $U(\sigma)$ was introduced in ref.~\citet{Boguta} to fix the incompressibility, is given by:
\begin{equation}
U(\sigma) =  \frac{\kappa M_N(g_{\sigma} \sigma)^3}{3} + \frac{\lambda(g_{\sigma}\sigma)^4}{4} \label{s2} .
\end{equation} 

The Pauli matrices are denoted by $\vec{\tau}$,  the antisymmetric mesonic field strength tensors are given by their usual expressions: $\Omega_{\mu\nu} = \partial_\mu \omega_\nu - \partial_\nu \omega_\mu$, and $P_{\mu\nu} = \partial_\mu \vec{\rho}_\nu - \partial_\nu \vec{\rho}_\mu -g_{\rho}(\vec{\rho}_\mu \times \vec{\rho}_\nu)$.   The $\gamma^\mu$ are the Dirac matrices and $\Lambda_{\omega\rho}$ is a non-linear isoscalar-isovector mixing coupling that provides a simple and efficient method of softening the symmetry energy without compromising the success of the model in reproducing well-determined ground-state observables~\citep{IUFSU}.  The detailed calculation of the EOS in the mean field approximation can be found in refs.~\citet{Glenbook,Serot_1992} and the references therein.

\begin{table}[h]
\begin{center}
\begin{tabular}{cc|cccc}
\toprule
  & Parameters & &  Constraints  & This model  \\
\toprule
 $\left(g_{\sigma}/{m_s}\right)^2$ & $12.108 \, \mathrm{fm}^2$ &$n_0 (\mathrm{fm}^{-3})$ & 0.148 - 0.170 & 0.156 \\
$\left(g_{\omega}/{m_v}\right)^2$  & $7.132 \, \mathrm{fm}^2$ & $M^{*}/M$ & 0.6 - 0.8 & 0.69  \\
  $\kappa$ & 0.004138 & $K \mathrm{(MeV)}$ & 220 - 260             &  256  \\
$\lambda$ &  -0.00390  & $S_0 \mathrm{(MeV)}$  & 30 - 35 &  31.7  \\
- &  - & $B/A \mathrm{(MeV)}$  & 15.8 - 16.5  & 16.2 \\
\toprule
\end{tabular}
\caption{Model parameters used in this study and their predictions for symmetric nuclear matter at saturation density. The parametrization was taken from ref.~\citet{Lopes2023PRD2}, and the  phenomenological constraints were taken from refs.~\cite{Dutra2014, Micaela2017}.}
\label{TL1}
\end{center}
\end{table}

\begin{center}
\begin{table}[ht]
\begin{center}
\begin{tabular}{cccccc}
\toprule
  $L$ (MeV) & $K_{sym}$ (MeV) & $(g_\rho/m_\rho)^2$ ($\mathrm{fm}^2$) & $(g_\delta/m_\delta)^2$ ($\mathrm{fm}^2$)  & $\Lambda_{\omega\rho}$  \\
\toprule
 44 &-58 &8.40 & 0  & 0.0515 \\
 76&-12 &4.90 & 0 & 0.0171        \\
100 & 121 &7.23 & 0.92 & 0   \\
116 &209 & 13.48 & 2.76 & 0   \\
\toprule
\end{tabular}
\caption{Model parameters selected to set the symmetry energy at $S_0$ = 31.7 MeV taken from ref.~\cite{Lopes2023PRD2}.} 
\label{T2}
\end{center}
\end{table}
\end{center}

In this work, I follow ref.~\citet{Lopes2023PRD2} and use a slightly modified version of the L3$\omega\rho$ parametrization proposed in ref.~\citet{Lopes2022CTP}.
The model parameters and their predictions for symmetric nuclear matter are presented in Table \ref{TL1}. The five well-known nuclear constraints at saturation density discussed above are taken from two extensive review articles:  \citet{Dutra2014, Micaela2017}, and are also included in Table \ref{TL1}. The parameters $(g_\rho/m_\rho)^2$, $(g_\delta/m_\delta)^2$, and $\Lambda_{\omega\rho}$ are chosen in order to fix the symmetry energy at the saturation point $S_0$ = 31.7 MeV while varying the slope $L$. The values of these parameters are reported in Table~\ref{T2}. As the parameterizations presented in this work are the same as presented in ref.~\citet{Lopes2023PRD2}, additional discussion about the symmetry energy and its slope, as well complementary results for non-rotating neutron stars can be found in ref.~\citet{Lopes2023PRD2}. Furthermore, in Table \ref{T2} we also calculate the second derivative of the symmetry energy~\citep{Zhang_2018}:

\begin{equation}
K_{sym} =  9n_0^2 \bigg ( \frac{\partial^2 S}{\partial n^2} \bigg ) \bigg |_{n_0}.
\end{equation}
This quantity  was not calculated in previous works for the present parameterizations. As can be seen, $K_{sym}$ is negative for lower values of $L$, and positive for higher ones. From the field theory point of view, $K_{sym}$ is negative for models that present non-linear $\omega-\rho$ coupling and positive for models that incorpore the scalar-isovector $\delta$ meson.

\section{Structural Equations and results}

\subsection{Static Limit}

For non-rotating neutron stars, the equations of the hydrostatic equilibrium are the so-called 
Tolman-Oppenheimer-Volkof (TOV) equations~\citep{TOV2}, which reads:

\begin{eqnarray}
 \frac{dp}{dr} = \frac{-M(r)\epsilon (r)}{r^{2}} \bigg [ 1 + \frac{p(r}{\epsilon(r)} \bigg ] \quad 
 \nonumber 
  \bigg  [ 1 + \frac{4\pi p(r)r^3}{M(r)} \bigg ] \bigg [ 1 - \frac{2M(r)}{r} \bigg ]^{-1}, \label{TOVS1}  \\
 \frac{dM}{dr} =  4\pi r^2 \epsilon(r) , \label{TOVS2} 
\end{eqnarray}
where $\epsilon$ is the energy density and $p$ is the pressure, which must be obtained from an equation of state (EOS), $M(r)$ is the mass enclosed in radius $r$. Moreover, here I use G = c = 1.

\begin{figure}[h]
  \begin{centering}
\includegraphics[width=0.333\textwidth,angle=270]{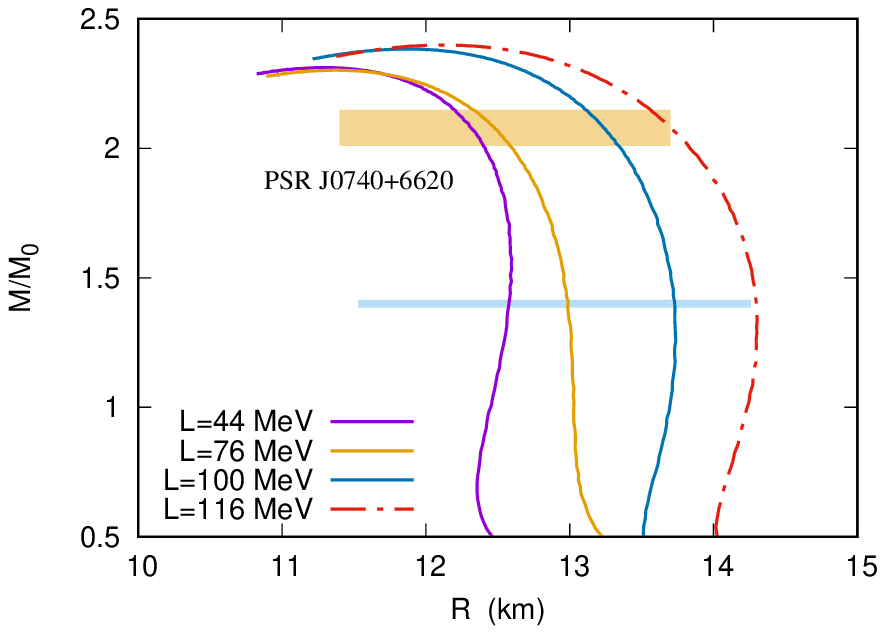} \\
\caption{Mass-radius relation for different values of $L$ for non-rotating neutron stars. The hatched areas are the constraints discussed in the text.} \label{Ftov}
\end{centering}
\end{figure}

In Fig.~\ref{Ftov}, we display the mass-radius relation coming from the TOV solution for different values of the symmetry energy slope.   For all parameterizations, we use the BPS EOS~\citep{BPS} for the neutron star's outer crust and the BBP EOS~\citep{BBP} for the inner crust.  We use the BPS+BBP EoS up to the density of 0.0089 fm$^{-3}$ for all values of $L$, and from this point on, we use the QHD EoS, as suggested in ref.~\citet{Glenbook}.  Imposing  $p_{crust} = p_{core}$ at the crust-core transition, I found that the core EoS begins around 0.03 fm$^{-3}$ for $L =  44$ MeV, up to around 0.05 fm$^{-3}$ for $L = 116$ MeV. In ref.~\cite{Fortin2016}, the authors compare the BPS+BBP crust EoS with a unified EoS. They show that for the canonical star, there is a variation in the radius of $60 \mathrm{m} <~R_{1.4}~< 150 \mathrm{m}$. For a radius of 13 km, this implies an uncertainty around 1$\%$.  It is also worth pointing out that unified EOS predict that the crust-core phase transition happens at higher densities, around 0.08 fm$^{-3}$ (see ref.~\citet{Carreau2019,Grams2022} to additional discussion).  The neutron stars' core EOS derived from the QHD are chemically stable and electrically neutral. We also discuss a couple of observational constraints coming from the NICER X-ray telescope. The first, and maybe the most important one, is the PSR J0740+6620, whose mass and radius lie in the range of $M = 2.08\pm 0.07~M_\odot$, and 11.41 km $<~R~<$ 13.70 km, respectively~\citep{Riley2021}.
The other constraint is related to the radius of the canonical $M = 1.4~M_\odot$ star.
Two NICER results constrain the radius of the canonical star between  11.52 km $<~R_{1.4}~<$ 13.85 km~\citep{Riley_2019} and between 11.96 and 14.26 km~\citep{Miller_2019}. Here we use the union set of both constraints as a more conservative approach. Explicitly, we use 11.52 km $<~R_{1.4}~<$ 14.26 km as a constraint.  The correlation between $L$ and the radii of neutron stars (i.e., that models with larger $L$ have larger radii) has been well-established in the literature~\citep{Rafa2011,dex19jpg,Lopes2023PRD2}. Moreover, as can be seen, the increase of the slope causes a small increase in the maximum mass.

In Newtonian formalism, the mass shedding limit, which is the maximum angular velocity a gravitational bounded object can rotate is given by~\citep{Hartle_1967,Hartle_1968}:

\begin{equation}
\Omega_K = \sqrt{\frac{M}{R^3}}.  
\end{equation}

As the gravity in general relativity is stronger than his Newtonian counterpart, one can expect that a relativistic star can rotate even faster than the Newtonian limit. However, it was shown in ref.~\citet{Weber1992}, that it is the opposite. The Kepler frequency is actually lower due to the drag of the inertial frame. A more precise expression for the Kepler frequency is given by the empirical formulae:

\begin{equation}
\Omega_K = 0.65 \sqrt{\frac{M}{R^3}}.  \label{kf}
\end{equation}

\begin{figure}[h]
  \begin{centering}
\includegraphics[width=0.333\textwidth,angle=270]{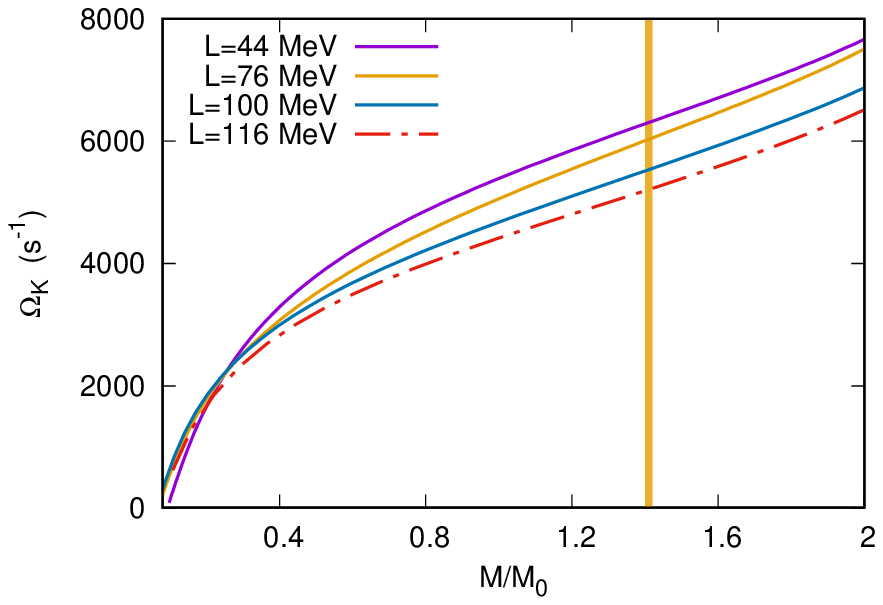} \\
\caption{Kepler frequency in function of the neutron stars mass for different values of $L$. The vertical line corresponds to the canonical $M = 1.4M_\odot$ star.} \label{Fkep}
\end{centering}
\end{figure}

As pointed out in ref.~\citet{Glenbook,Weber1992}, the great advantage of eq.~\ref{kf} is the fact that the relativistic Kepler frequency of a neutron star rotating at its mass shedding limit can be
estimated from the mass and radius of the corresponding non-rotating limiting-mass star. In Fig~\ref{Fkep}, we display the Kepler limit in function of the neutron stars' mass for different values of $L$. Moreover, in Tab~\ref{TK}, we display the inferior mass limit for four different values of $\Omega$: 1500 $s^-1$,  3000 $s^-1$, 4500 $s^-1$, and 5000 $s^-1$.
We can notice that as we increase the angular velocity, we also increase the inferior limit of the neutron star mass ($M_{lim}$), where, in the Hartle-Throne approximation, only the results for neutron stars with $M~>M_{limt}$ are valid. Moreover, larger values of $L$ present higher values of $M_{lim}$ for a fixed $\Omega$. This was expected, due to the relation between the radii and the slope. For $L$ = 116  MeV, we see that the inferior limit is close to the canonical mass for $\Omega = 5000~s^{-1}$. An interesting feature is the fact that for $\Omega = 1500~s^{-1}$, the lower values of $L$ present the higher value of $M_{lim}$. The reason is for such low angular velocity, we have a very low neutron star mass with a low central density. And, as shown in Fig. 2 in ref.~\citet{Lopes2023PRD2}, for low densities, the EOS is stiffer for low values of $L$. Then it is reversed for high densities.

\begin{center}
\begin{table}[bt]
\begin{center}
\begin{tabular}{ccc}
\toprule
  $L$ (MeV) & $\Omega~(s^{-1})$) & $M_{lim}/M_\odot$  \\
\toprule
 44 & 1500 & 0.19   \\
 44 & 3000 & 0.35    \\
 44 & 4500 & 0.68   \\
 44 & 5000 & 0.87   \\
\hline
 76& 1500 & 0.17       \\
 76& 3000 & 0.39        \\
 76& 4500 & 0.79        \\
 76& 5000 & 0.98       \\
 \hline
100 &  1500 & 0.16   \\
100 &  3000 & 0.41    \\
100 &  4500 & 0.92  \\
100 &  5000 & 1.16 \\
\hline 
116 &  1500 & 0.16  \\
116 &  3000 & 0.44   \\
116 &  4500 & 1.04   \\
116 &  5000 & 1.31   \\
\toprule
\end{tabular}
\caption{Inferior limit of the neutron stars' masses for a fixed $\Omega$.} 
\label{TK}
\end{center}
\end{table}
\end{center}

\subsection{Rotating neutron stars}

The detailed formalism about slowly rotating neutron stars can be found in the Hartle and Thorne original papers~\citep{Hartle_1967,Hartle_1968,Hartle_1973}. Additional discussion can also be found in ref.~\citet{Glenbook,Weber1992,Chu2000,Pattersons2021}. Here we present only a small discussion and the necessary equations to build a numerical code. This is especially useful to help readers who are not familiar with the Hartle-Thorne (HT) formalism. The metric of a non-rotating, spherical symmetric star has the Schwarzschild form:

\begin{eqnarray}
ds^2 =  -e^{2\varphi} dt^2 +\bigg [ 1 - \frac{2M(r)}{r} \bigg ]^{-1}dr^2 
+ r^2 (d\theta^2 + \sin^2\theta d\phi^2), \label{nrm}
\end{eqnarray}
where:

\begin{equation}
\frac{d\varphi}{dr} = \frac{M(r) + 4\pi r^3p(r)}{r - 2M(r)}. \label{dphi}
\end{equation}

Therefore, we can write the metric for a slowly rotating star as a perturbed Schwarzschield metric:
\begin{eqnarray}
ds^2 = -e^{2\varphi}(1 + 2h(r,\theta))dt^2 
+ \bigg [1 + \frac{2m(r,\theta)}{r - 2M(r)} \bigg ] \bigg [ 1 - \frac{2M(r)}{r} \bigg ]^{-1}dr^2 
+ r^2(1 + 2k(r,\theta))[d\theta^2 + \sin^2\theta(d\phi - \omega dt)^2] , \label{rm}
\end{eqnarray}
where the perturbative terms can be expanded up to the second-order:

\begin{eqnarray}
 h(r,\theta) = h_0(r) + h_2(r)P_2(\cos\theta) ,  \nonumber \\
m(r,\theta) = m_0(r) + m_2(r)P_2(\cos\theta), \nonumber \\
 k(r,\theta) = k_0(r) + k_2(r)P_2(\cos\theta) , \label{pert}
\end{eqnarray}
where $P_2(\cos\theta)$ is the Legendre polynomial of order 2. Due to the symmetry $k_0(r)$ = 0. Moreover, as will become clear at the end of the calculations, the terms $h_0(r)$, and $m_2(r)$  do not contribute to the mass correction nor to the star deformation in our approximation, and will not be taken into account any further (they expression are nevertheless presented in ref.~\citet{Hartle_1968}). We also define $\nu_2~\equiv~h_2 +k_2$ as suggested in ref.~\citet{Hartle_1967}. The parameter $\omega = \omega(r) = d\phi/dt$ is the angular velocity acquired by a free-falling observer due to the drag of the inertial frame. The angular velocity of the star relative to the local inertial frame, $\bar{\omega} = \Omega - \omega$ is obtained by solving:

\begin{equation}
 \frac{1}{r^4}\frac{d}{dr}\bigg ( r^4 j \frac{d\bar{\omega}}{dr} \bigg ) + \frac{4}{r}\frac{dj}{dr}\bar{\omega} = 0, \label{omegabar}
\end{equation}
with

\begin{equation}
j =  j(r) = e^{-\varphi} \bigg [1 - \frac{2M(r)}{r} \bigg ]^{1/2}.
\end{equation}

The angular momentum ($J$) of a spherical symmetrical star with radius $R$ is given by:
\begin{equation}
  J = \frac{1}{6}R^4\bigg (\frac{d\bar{\omega}}{dr} \bigg ) \bigg |_{r =R}  
\end{equation}

Now, if $p$,  $\epsilon$, and $n$ are the pressure, the energy density, and the number density at some given ($r, \theta)$ for a static star, the correspondent values of the pressure, energy density, and number density at the same given ($r, \theta)$  that is momentarily moving with the fluid in a rotating star are  $p + \Delta p$,  $\epsilon +\Delta \epsilon$, and $n + \Delta n$. The perturbation terms can also be expanded up to second-order:

\begin{eqnarray}
  \Delta p = (\epsilon +p)[p_0^* + p_2^*P_2(\cos(\theta)] , \nonumber \\
  \Delta \epsilon = (\epsilon +p)[p_0^* + p_2^*P_2(\cos(\theta)](d\epsilon/dp) , \nonumber \\ 
  \Delta n = (\epsilon +p)[p_0^* + p_2^*P_2(\cos(\theta)](dn/dp) , \label{pertub2}
\end{eqnarray}
where, $p_0^*$ and $p_2^*$ are dimensionless functions of $r$. Now, the perturbative terms, $m_0$, $p_0^*$, $\nu_2$, $h_2$,  and $p_2^*$ are obtained by applying Einstein's field equations. We obtain:

\begin{equation}
\frac{dm_0}{dr} = 4\pi r^2 \frac{d\epsilon}{dp}(\epsilon + p)p_0^* + \frac{1}{12}j^2r^4 \bigg (\frac{d \bar{\omega}}{dr} \bigg )^2  
 - \frac{1}{3} r^3 \frac{dj^2}{dr} \bar{\omega}^2, \label{dmo}
\end{equation}

\begin{equation}
\frac{dp_0^*}{dr} = - \frac{m_0(1 +8\pi r^2 p)}{(r - 2M(r))^2} - \frac{p_0^* (\epsilon +p)4\pi r^2 }{(r -2M(r))}  
+ \frac{1}{12} \frac{r^4j^2}{(r - 2M(r))} \bigg ( \frac{d\bar{\omega}}{dr} \bigg )^2 + \frac{1}{3} \frac{d}{dr} \bigg ( \frac{r^3 j^2 \bar{\omega}^2}{r - 2M(r)} \bigg ), \label{dp0}
\end{equation}

\begin{equation}
\frac{d\nu_2}{dr} = -2h_2 \bigg ( \frac{d\varphi}{dr} \bigg ) + \bigg ( \frac{1}{r} +\frac{d\varphi}{dr} \bigg ) 
 \bigg [- \frac{1}{3} r^3 \frac{dj^2}{dr} \bar{\omega}^2   + \frac{1}{6} j^2r^4 \bigg (\frac{d \bar{\omega}}{dr} \bigg )^2  \bigg ] , \label{dnu2}
\end{equation}

\begin{eqnarray}
\frac{dh_2}{dr} = -2h_2 \bigg ( \frac{d\varphi}{dr} \bigg )  + \frac{r}{r - 2M(r)} \bigg (2 \frac{d\varphi}{dr} \bigg )^{-1}  
\bigg [ 8\pi (\epsilon + p) - \frac{4M(r)}{r^3} \bigg ]h_2  
- \frac{4\nu_2}{r(r -2M(r))} \bigg (2 \frac{d\varphi}{dr} \bigg )^{-1} \nonumber \\
+ \frac{1}{6} \bigg [ \frac{d\varphi}{dr}r - \frac{1}{r -2M(r)} \bigg (2 \frac{d\varphi}{dr} \bigg )^{-1}\bigg ]r^3j^2 \bigg (\frac{d \bar{\omega}}{dr} \bigg )^2  
- \frac{1}{3} \bigg [ \frac{d\varphi}{dr}r - \frac{1}{r -2M(r)} \bigg (2 \frac{d\varphi}{dr} \bigg )^{-1}\bigg ]r^3 \bigg (\frac{dj^2}{dr} \bigg ) \bar{\omega}^2 , \label{dh2}
\end{eqnarray}  
with the boundary conditions: $m_0(0) = p_0^*(0) = h_2(0) = \nu_2(0) = 0$. Finally, we have:

\begin{equation}
 p_2^* = -h_2 - \frac{1}{3}r^2 e^{-2\varphi} \bar{\omega}^2,
\end{equation}
which also implies $p_2^*(0)$ = 0. The mass correction is given by:

\begin{equation}
  \delta M = m_0 + \frac{J^2}{R^3},  
\end{equation}
and the deformation of the star is represented by:

\begin{equation}
 \delta r(r,\theta) =     \xi_0(r) + \xi_2(r)P_2(\cos\theta),
\end{equation}
where

\begin{eqnarray}
  \xi_0 = -p_0^*(\epsilon + p)\bigg (\frac{dp}{dr} \bigg )^{-1}, \nonumber \\
  \xi_2 = -p_2^*(\epsilon + p)\bigg (\frac{dp}{dr} \bigg )^{-1} \label{xis} .
  \end{eqnarray}

Therefore, if a non-rotating spherical symmetric neutron star with a central density $n_c$ presents a mass $M$ and a radius $R$, for the same central density, a slowly rotating neutron star will be a spheroid with mass $M + \delta M$ and a radius $R(\theta) =  R +\delta R(R,\theta)$. Nevertheless, they are not the same star, as they present different baryon numbers, as pointed out in  ref.~\citet{Glenbook}. Finally, we can express the polar ($R_p$) and equator $(R_e)$ radii as~\citep{Pattersons2021}:

\begin{eqnarray}
 R_{po} =  R + \xi_0 + \xi_2, \nonumber \\
 R_{eq} =  R + \xi_0 - \frac{1}{2}\xi_2 , \nonumber \label{radii}
\end{eqnarray}
this allows us to define the eccentricity:

\begin{equation}
  e =  \sqrt{ 1 - \bigg (\frac{R_{po}}{R_{eq}} \bigg )^2}. \label{ecc}  
\end{equation}

As the measurement of the angular velocity is an easy task,  once it is related to the period: $\Omega = 2\pi/T$, where $T$ is the period, and the eccentricity does not depend on the absolute values of the radii, this quantity can play a very special role to fix the symmetry energy slope.

\begin{figure*}[t]
\begin{tabular}{cc}
\centering 
\includegraphics[scale=.51, angle=270]{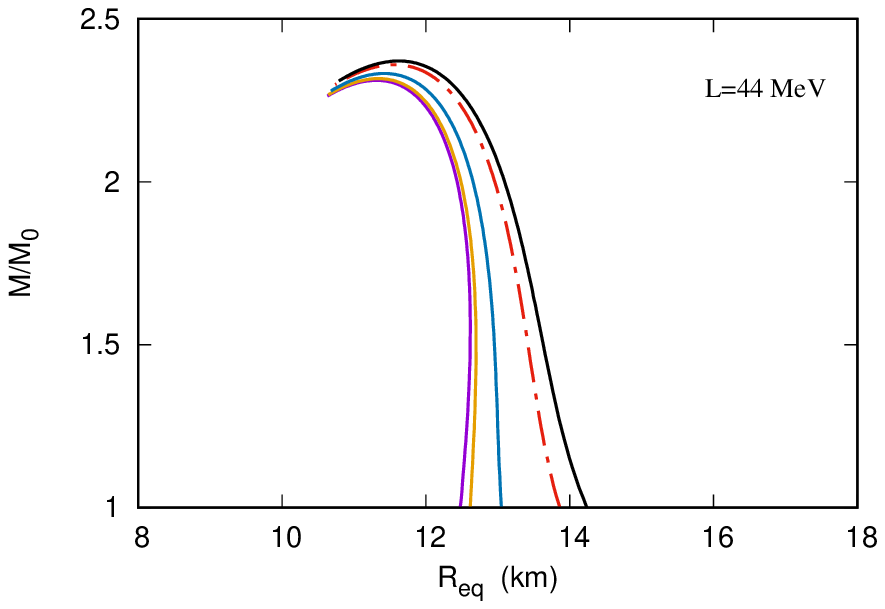} &
\includegraphics[scale=.51, angle=270]{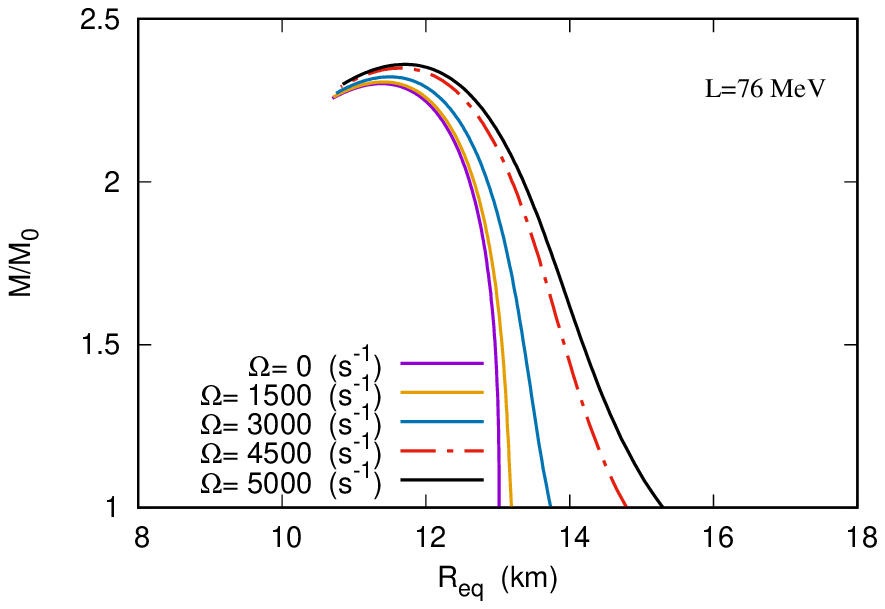}\\
\includegraphics[scale=.51, angle=270]{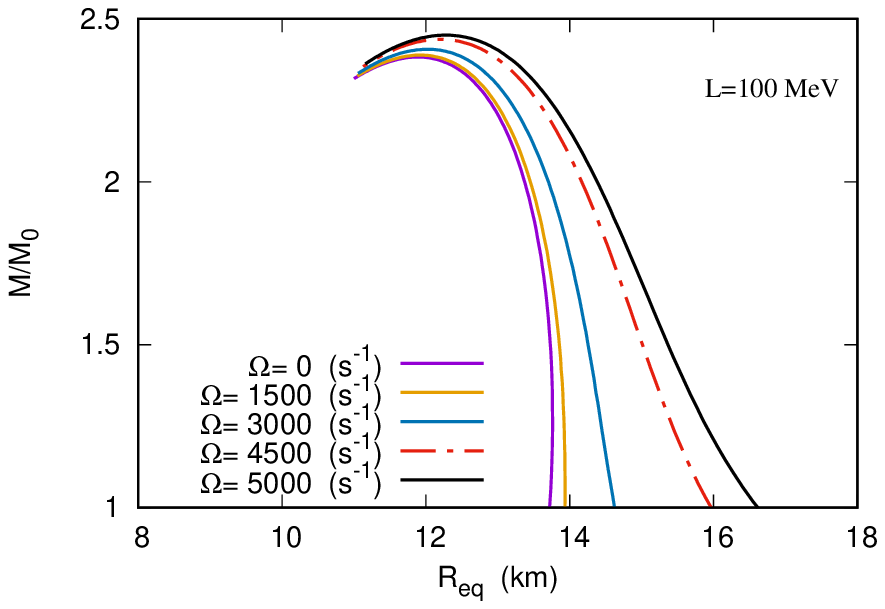} &
\includegraphics[scale=.51, angle=270]{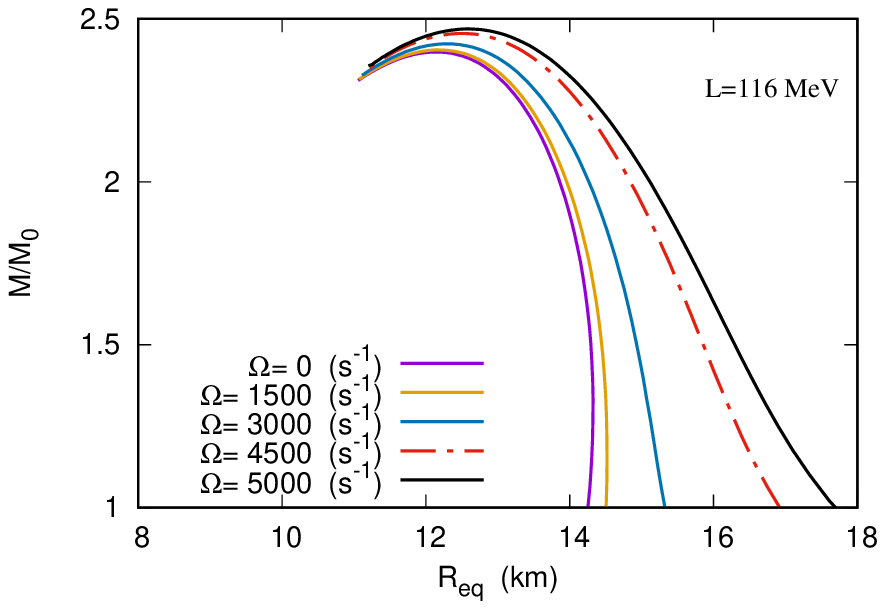}\\
\end{tabular}
\caption{(Color online) Equator radius as a function of the stellar mass for different values of $L$ and different values of $\Omega$. } \label{F3}
\end{figure*}

Once I established the basic equations of the  Hartle-Thorne formalism, I now solve these coupled equations for four different values of $L$: 41, 76, 100, and 116 MeV.  And for each value of $L$ we study four different values of angular velocities:  $\Omega$: 1500 $s^-1$,  3000 $s^-1$, 4500 $s^-1$, and 5000 $s^-1$.
The focus here is to investigate how different values of the slope change the main properties of neutron stars, I especially focus on the maxim mass, the correspondent central density, and the variation of the radius of some fixed mass. The first one is the canonical 1.4$M_\odot$ and the second is the  2.01 $M_\odot$ star, which is not only the most probable mass value of PSR J0348+0432~\citep{Antoniadis_2013} but also the lower limit of PSR J0740+6620, whose gravitational mass is 2.08 $\pm$ 0.07 $M_\odot$~\citep{Miller_2021,Riley2021}.

\begin{figure*}[h!]
\begin{tabular}{cc}
\centering 
\includegraphics[scale=.51, angle=270]{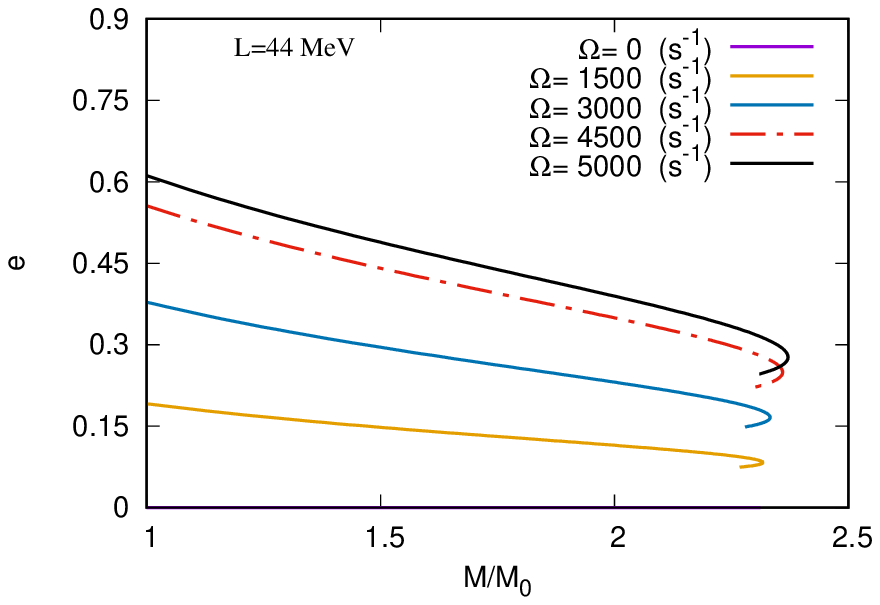} &
\includegraphics[scale=.51, angle=270]{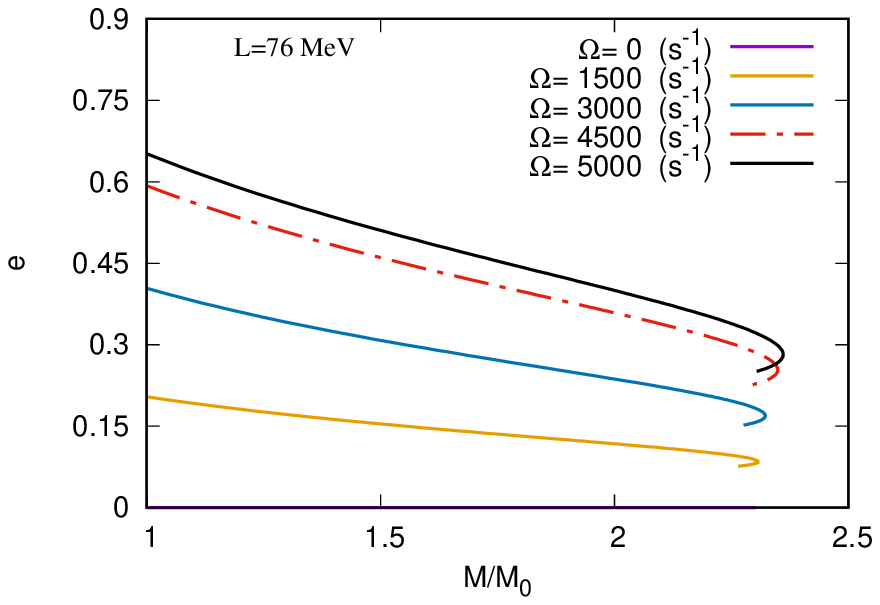}\\
\includegraphics[scale=.51, angle=270]{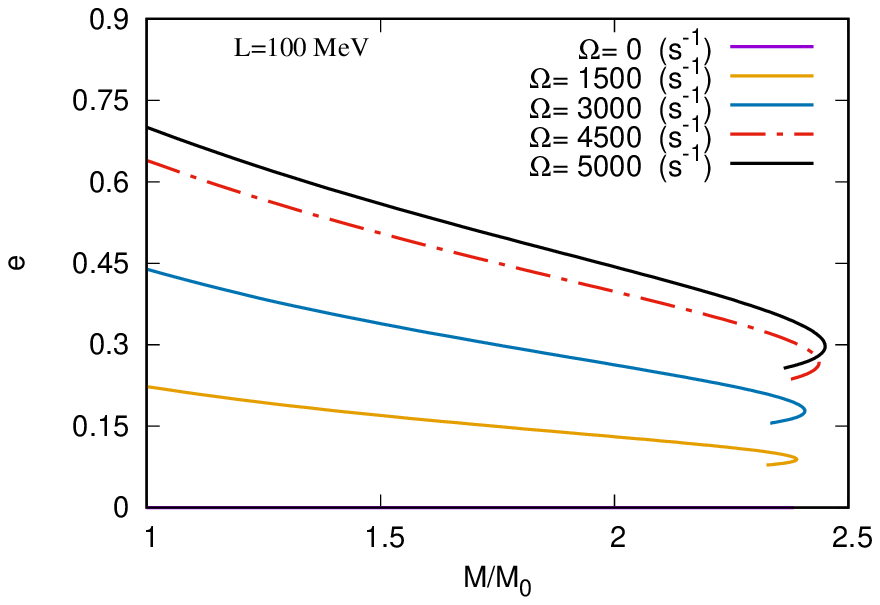} &
\includegraphics[scale=.51, angle=270]{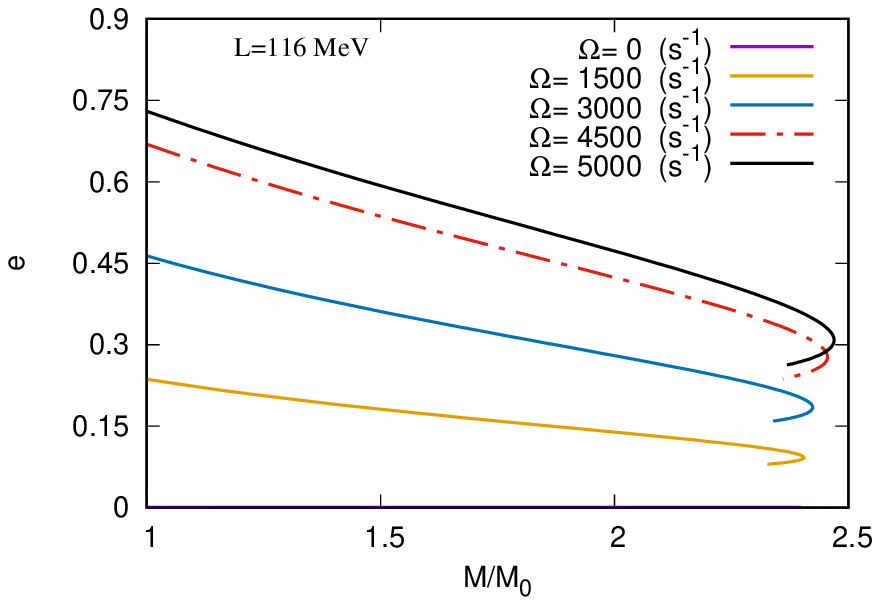}\\
\end{tabular}
\caption{Eccentricity of rotating neutron stars as a function of the stellar mass for different values of $L$ and different values of $\Omega$.} \label{F4}
\end{figure*}

In Fig.~\ref{F3} I display the equator radius in function of the stellar mass for rotating neutron stars, for different values of $L$ and $\Omega$, while in  fig~\ref{F4} I show the eccentricity of the star in function of the mass for the same $L$ and $\Omega$. From a qualitative point of view, the effect of rotation is the same for all values of $L$. As we increase the angular velocity, we also increase the equator radius for all values of a fixed $M$. We also increase the maximum mass.
The same is true of the eccentricity.

From the quantitative point of view, we see that the equator radius of the canonical star increases from 12.58 km for a non-rotating neutron star up to 13.72 km for a rotating neutron star with $\Omega = 5000~(s^{-1})$ in the case of $L = 44$ MeV, which represents an increase of 1.14 km or  $9.1\%$.
For the case of $L = 116$ MeV, we see that the equator radius of the canonical star increases from 14.30 km up to 16.51 km. An increase of 2.21 km or 15$\%$. Therefore, not only the absolute values of $R_{eq}$ are dependent on the slope but also the relative ones. This means that for the same angular velocity, the star will appear flatter if the slope is higher. This can easily be seem from the eccentricity. While for $L$ = 44 MeV we have $e = 0.51$ for $\Omega = 5000~(s^{-1})$, for $L$ = 116 MeV we have   $e = 0.61$ for the same angular velocity. The same is true for $M = 2.01~M_\odot$ but with more modest results. The equator radius for a 2.01 solar mass star increases from 12.40 km up to 13.07 km for $L = 44$ MeV and from 13.82 km up to 15.03 km for $L$ = 166 MeV. Both at $\Omega = 5000~s^{-1}$. This implies an increase of 0.67 km or $5.4\%$ for $L$ = 44 MeV and 1.48 km or $11\%$ for $L$ = 116 MeV. The eccentricities are  0.39 and 0.46 for $L$ = 44 and 116 MeV respectively.

I now analyze the variation of the maximum mass and the correspondent central density for $\Omega = 5000~s^{-1}$. For $L$ = 44 the maximum mass increase from 2.31 $M_\odot$ up to 2.37 $M_\odot$. A change of 0.06 $M_\odot$ or $2.6\%$. The central density drops from 0.939 fm$^{-3}$ to 0.899 fm$^{-3}$, or a decrease of $4.3\%$. On the other hand, the maximum mass increases from 2.39 $M_\odot$ up to 2.47 $M_\odot$, an increment of $3.3\%$,  while the central density decreases from 0.872 fm$^{-3}$ to 0.823 fm$^{-3}$, which represent a decrease of $5.6\%$. 
All relevant quantities are summarized in Tab.~\ref{T4}. The percentage is expressed with two significant digits.  It is also important to emphasize that these results will be slightly different within a unified EOS.

For a fixed mass value and $\Omega$, the higher the symmetry energy, the higher will be the eccentricity. This fact can be very important in a future analysis of the 33 recently discovered millisecond pulsars~\citep{Smith2023}. As the measurement of the angular velocity is easy, and the eccentricity does not depend on the absolute value of the neutron stars' radii, its measure is a powerful tool, which combined with other techniques can improve the constraints of the symmetry energy slope.

\begin{widetext}
\begin{center}
\begin{table}[ht]
\begin{center}
\scriptsize
\begin{tabular}{cc|cccccccccccccc}
\hline
  $L (\mathrm{MeV})$ & $\Omega~(s^{-1})$ &$M_{\mathrm{max}} (M_\odot)$ &  $\Delta M$ & $n_c$ (fm$^{-3}$) & $\Delta n_c$  &$R_{(1.4)}$ (km) & $\Delta$  (km) & $\Delta~(\%)$ & $e_{1.4}$  & $R_{(2.01)}$ (km) & $\Delta$ (km) & $\Delta~(\%)$ & $e_{2.01}$  \\
\toprule
 44 & 0 & 2.31 &  - &0.939 & - & 12.58 & - & - & 0.00 &  12.40 & - & - & 0.00 \\
 44 &1500 & 2.32 & 0.43$\%$ &0.933 &-0.63$\%$ &12.69 & 0.11 & 0.87$\%$ & 0.15 &  12.46 & 0.06 & 0.48$\%$ & 0.11 \\
44 &3000 & 2.33 &0.86$\%$ & 0.923& -1.7 $\%$  &12.97 & 0.39 & 3.1$\%$ & 0.31 &  12.64 & 0.24 &1.9$\%$ & 0.23 \\ 
44 &4500 & 2.36 & 2.1$\%$ &0.909 & -3.2$\%$ & 13.48 &0.90 & 7.1$\%$ & 0.46 &  12.94 &0.54 & 4.3$\%$ & 0.35 \\
44 &5000 & 2.37 & 2.6$\%$ & 0.899 &-4.3$\%$  &13.72 &1.14  & 9.1$\%$ & 0.51 &  13.07 & 0.67 & 5.4$\%$ & 0.39 \\
 \hline
 76& 0 &2.30 & -& 0.938 & - & 12.99 & - &- & 0.00   &  12.59 & - &  - & 0.00  \\
 76&1500 &2.31 & 0.43$\%$ & 0.934& -0.43$\%$ &13.10 &0.11 & 0.84$\%$ & 0.16   &  12.65 &0.06 &  0.47$\%$ & 0.12  \\
 76&3000 &2.32 & 0.87$\%$ &0.924 & -1.5$\%$ &13.42 &0.43 & 3.2$\%$ & 0.32   &  12.83 & 0.24 &  1.9$\%$  & 0.23  \\
 76&4500&2.35 & 2.2$\%$ & 0.908 & -3.2$\%$ &14.06 & 1.07 & 8.2$\%$ & 0.48   &  13.18 & 0.59 &  4.7$\%$ & 0.36  \\
 76&5000 &2.36 & 2.6$\%$ &0.898 & -4.3$\%$ &14.35 & 1.36 & 10$\%$ & 0.53   &  13.34 & 0.75& 6.0$\%$ & 0.40  \\
\hline
100 & 0 &2.37 & -  & 0.892 & - &13.73 & - & -  &  0.00 &  13.34 & - & - &  0.00  \\
100 & 1500 &2.38& 0.42$\%$ & 0.885&-0.78$\%$ &13.89&0.16 & 1.1$\%$ & 0.18 &  13.42 &0.08& 0.60$\%$ &  0.13 \\
100 & 3000 &2.40 & 1.3$\%$ & 0.872& -2.2$\%$ &14.33 &0.60 & 4.4$\%$ & 0.36 & 13.67 &0.34 & 2.5$\%$ &  0.26  \\
100 & 4500 &2.44 &2.9$\%$ &0.855 &-4.1$\%$ &15.18 &1.45 & 11$\%$ & 0.53 &  14.14 &0.80&   6.0$\%$  &  0.40  \\
100 & 5000 &2.45 & 3.3$\%$&0.847 & -5.0$\%$ &15.55 &1.83 & 13$\%$ & 0.58 &  14.34 &1.00& 7.5$\%$ &  0.44  \\
\hline 
116 & 0 &2.39 & - &0.872 & - &14.30  &- & - & 0.00 & 13.82 & - &  - & 0.00  \\
116 & 1500 &2.40 & 0.41$\%$ &0.866 & -$0.68\%$  &14.48 & 0.18 & $1.3\%$ & 0.19 &  13.94 &0.12 & $0.86\%$ & 0.14  \\
116 & 3000 &2.42 & 1.3$\%$ &0.852 & -$2.3\%$ &15.01 &0.71 & $5.0\%$ & 0.38 &  14.25 &0.43 &  3.1$\%$ & 0.28  \\
116 & 4500 &2.46 &$2.9\%$ & 0.831& -$4.7\%$ &16.05 &1.75 & 12$\%$ & 0.56 & 14.82 &1.00 &  $7.2\%$ & 0.42  \\
116 & 5000 &2.47 & 3.3$\%$ &0.823 &  -5.6$\%$ &16.51 &2.21 & $15$\% & 0.61 &  15.03 &1.21&  11$\%$ & 0.46 \\
\hline
\end{tabular}
\caption{Neutron stars' main properties for different values of $L$ and $\Omega$ } \label{T4}
\end{center}
\end{table}
\end{center}
\end{widetext}

\section{Conclusion}

In this work, I study the effects of the symmetry energy slope on rotating neutron stars. For four different values of $L$, I choose four values of $\Omega$, and see how some macroscopic properties are affected: the maximum mass, the central number density of the maximum mass, the equator radius, and the eccentricity of the canonical 1.4$M_\odot$ star and the 2.01 $M_\odot$ star.
We can see that the rotation affects all the neutron stars' properties. Moreover, the higher the slope, the higher the perturbation induced by the rotation, in both: absolute and relative terms.
For a fixed value of $\Omega$, we see that the effects are small in the maximum mass, are more significant at the 2.01 $M_\odot$ star, but strongly affect the canonical star.
Furthermore, as the eccentricity does not depend on the absolute value of the star's radius, and the measurement of angular velocity is not a hard task, this quantity can potentially help to constraint the symmetry slope, especially in light of the new 33 recently discovered millisecond pulsars. A value as high as 0.61 implies a spheroid star, whose equator radius is around 17$\%$ larger than the polar one.

\section*{acknowledgments}
I thank Anna Campoy for making her Fortran code publicly available  (see ref.~\cite{Campoy2019}) The code used in this work is based on hers.

%


\bibliography{rot}{}
\bibliographystyle{aasjournal}



\end{document}